# Versatile electronic states in epitaxial thin films of (Sn-Pb-In)Te: from topological crystalline insulator and polar semimetal to superconductor

Ryutaro Yoshimi[1*], Makoto Masuko[2], Naoki Ogawa[1], Minoru Kawamura[1], Atsushi Tsukazaki[3], Kei S. Takahashi[1], Masashi Kawasaki[1,2], and Yoshinori Tokura[1,2,4]

[1] RIKEN Center for Emergent Matter Science (CEMS), Wako 351-0198, Japan.

[2] Department of Applied Physics and Quantum-Phase Electronics Center (QPEC), University of Tokyo, Tokyo 113-8656, Japan

[3] Institute for Materials Research, Tohoku University, Sendai 980-8577, Japan

[4] Tokyo College, University of Tokyo, Tokyo 113-8656, Japan

\* Corresponding author: ryutaro.yoshimi@riken.jp

## Abstract

Epitaxial thin films of $(Sn_xPb_{1-x})_{1-y}In_yTe$ were successfully grown by molecular-beam-epitaxy (MBE) in a broad range of compositions ($0 \leq x \leq 1, 0 \leq y \leq 0.23$). We investigated electronic phases of the films by the measurements of electrical transport and optical second harmonic generation. In this system, one can control the inversion of band gap, the electric polarization that breaks the inversion symmetry, and the Fermi level position by tuning the Pb/Sn ratio and In composition. A plethora of topological electronic phases are expected to emerge, such as topological crystalline insulator, topological semimetal, and superconductivity. For the samples with large Sn compositions ($x > 0.5$), hole density increases with In composition ($y$), which results in the appearance of superconductivity. On the other hand, for those with small Sn compositions ($x < 0.5$), increase in In composition reduces the hole density and changes the carrier type from p-type to n-type. In a narrow region centered at $(x, y) = (0.16, 0.07)$ where the n-type carriers are slightly doped, charge transport with high mobility exceeding 5,000 $cm^2V^{-1}s^{-1}$ shows up, representing the possible semimetal states. In those samples, the optical second harmonic generation measurement shows the breaking of inversion symmetry along the out-of-plane [111] direction, which ensures the presence of polar semimetal state. The thin films of $(Sn_xPb_{1-x})_{1-y}In_yTe$ materials systems with a variety of electronic states would become a promising materials platform for the exploration of novel quantum phenomena.

**Main text**

Recently chalcogenide compounds have attracted revived interest for novel physical phenomena due to their topological nature of electronic states. To name a few, $Bi_2Te_3$, $Sb_2Te_3$ and $Bi_2Se_3$ have been intensively studied as topological insulators with inverted band structure [1-4]; transition-metal dichalcogenides such as $WTe_2$ and $MoS_2$ have intriguing channels for valleytronics [5], superconductors [6], and quantum spin Hall effect [7]; Fe(Se,Te) is known as a superconductor and recently found to exhibit a feature of topological superconductivity with spin-helical surface states [8]. In addition to the remarkable physical properties in respective materials, combinations of these electronic states in the form of heterostructures would host further new exotic phenomena, including proximity effects of TI junctions with superconductivity [9-11] and magnetism [12,13]. Versatile electronic states in chemically-similar materials such as all telluride-based compounds are useful to design thin-film heterostructures and explore exotic quantum phenomena.

SnTe is one of rocksalt tellurides that shows a topological phase termed topological crystalline insulator (TCI). In TCI, two-dimensional topological surface states appear due to the inversion of bulk bands, and are protected by mirror reflection symmetry of the crystal structure [14-16]. The presence of surface states are experimentally verified by angle resolved photoemission spectroscopy [15,17]. The band inversion can be controlledby substitution of Sn with Pb, which enables us to explore the topological phase transition from TCI to trivial insulator [18]. Between the two phases, emergence of topological semimetals such as Weyl semimetals or nodal line semimetals by breaking inversion symmetry is theoretically predicted [19]. Experimentally, the existence of semimetal phase is suggested by the optical and transport measurements with the help of inherent ferroelectric instability of rocksalt crystal structure as well as external pressure [20-22]. Furthermore, the Fermi level can be modulated by doping In, which can realize not only high-mobility transport in the semimetal states but also superconductivity [23,24]. These novel electronic states in PbTe-SnTe system doped with In have been extensively studied in a bulk crystal form. In contrast, such an exploration of various electronic phases in a thin film form remains elusive except for Bi-doped (Sn,Pb)Te thin films [25], possibly due to the difficulty in precise tuning of chemical compositions. The thin films as well as the heterostructures, however, are indispensable for the study of quantum transport and device physics, in particular on the topological states of matter.

Here we investigate the electronic states in $(Sn_xPb_{1-x})_{1-y}In_yTe$ (SPIT) thin films by the measurements of electrical transport and optical second-harmonic-generation (SHG). By utilizing MBE thin film growth technique, we precisely tune the chemical composition and examine the effect of In-doping ($y \leq 0.23$) for SnTe-PbTe system over a full range of Sn/Pb composition ($0 \leq x \leq 1$). The superconductivity appears in the samples with $x \geq 0.63$ and $y \geq 0.17$, where the dopant of In acts as an accepter. On the other hand, the semimetal electronic states with remarkably high mobility transport is observed for samples in a narrow composition region at around $(x, y) \sim (0.16, 0.07)$ where the n-type carriers are slightly doped. The SHG measurement for high mobility samples detects the breaking of inversion symmetry characterized by the polarity along out-of-plane [111] direction, which supports the presence of polar semimetal phase. A rich variety of the electronic states would make the thin films of SPIT materials systems an ideal platform to explore novel quantum phenomena.

We grew SPIT thin films on InP(111)A substrates by MBE. The epi-ready substrates were annealed at 350 °C in a vacuum before the deposition at 400 °C of thin films. We inserted 2-nm thick SnTe buffer layers beneath the SPIT layer to stabilize the (111) orientation of SPIT thin films (Fig. 1(a)). The beam equivalent pressures of Sn and Te for the buffer layer were $P_{Sn} = 5.0 \times 10^{-6}$ Pa and $P_{Te} = 1 \times 10^{-4}$ Pa, respectively. For the SPIT layer, the equivalent pressures for cation elements (Sn, Pb and In) and anion element (Te) were set at $P_{Sn} + P_{Pb} + P_{In} = 1 \times 10^{-5}$ Pa and $P_{Te} = 1 \times 10^{-4}$ Pa, respectively. For example, $P_{Sn} = 4.0 \times 10^{-6}$ Pa, $P_{Pb} = 4.0 \times 10^{-6}$ Pa and $P_{In} = 2.0 \times 10^{-6}$ Pa for the nominal compositions of $(x, y) = (0.5, 0.25)$, where $x$ and $y$ respectively represent the compositions of Sn and In. Actual values of $x$ and $y$ in the films were calibrated by inductively coupled plasma mass spectroscopy. The calibrated $x$ value was larger than the nominal one by 0.1 ~ 0.2, suggesting that PbTe is more volatile than SnTe, whereas $y$ was almost the same as prescribed.

The growth duration of the SPIT layer was 30 minutes regardless of $x$. The thickness of SPIT was 30-40 nm which was precisely evaluated by X-ray reflectivity. The growth rate was evaluated to be 1.0-1.3 nm per minute, which depends on Pb/Sn flus ratio; it was fast when $P_{Sn} / P_{Pb}$ is large. As shown in the X-ray diffraction $2\theta$-$\omega$ scan (Fig. 1(b)) of $Sn_xPb_{1-x}Te$ films, sharp (111) and (222) diffraction peaks with clear Laue oscillations are commonly observed for all Sn composition $x$, indicating the high crystallinity of $Sn_xPb_{1-x}Te$ thin films with indiscernible secondary phases. The lattice constants of SnTe ($x = 1$) and PbTe ($x = 0$) evaluated from the (222) (Fig. 1(b)) diffraction peak show good agreements with those of bulk value (dashed lines in Fig. 1(c)). The $x$-dependence of the lattice constant follows the Vegard's law, suggesting that Pb is substitutionally replaced with Sn as intended. The reduction of the lattice constant is observed by In doping in the range of $y \leq 0.23$ (XRD data for $x = 0.63$ and 0.87 series are shown in Supplementary Materials).

The longitudinal and Hall resistivities were measured by the standard four-terminal method with use of Physical Properties Measurement System (PPMS, Quantum Design). The maximum magnetic field and the lowest temperature are 9 T and 1.8 K, respectively. Optical SHG was evaluated on the thin film samples mounted in a cryostat ($10 \text{ K} \leq T \leq 300 \text{ K}$) [26]. The 1.55 eV fundamental light (120 fs duration at 1 kHz repetition rate, 200 μW on a ~40 μmϕ spot) was incident at 45° off the normal of the film plane after its polarization set through a $\lambda/2$ plate. The generated second harmonic (SH) light at reflection geometry was directed to a Glan laser prism, color filters, and a monochromator, and detected with a photomultiplier tube. The signal was normalized by that of a reference potassium dihydrogen phosphate crystal, and accumulated more than 4000 times at each polarization configuration.

We begin with the results of electrical transport properties for In-free ($y = 0$) thin films of $Sn_xPb_{1-x}Te$. As shown in Fig. 1(d), longitudinal resistivity $\rho_{xx}$ exhibits a dramatic change depending on $x$, ranging over 4 orders of magnitude at $T = 2$ K. Both of the residual resistivity (Fig. 1(e)) and carrier density (Fig. 1(f)) at a temperature $T = 2$ K show monotonic $x$-dependence; smaller residual resistivity and larger carrier density with increasing $x$. Here, the carrier density is evaluated from the Hall resistivity at a magnetic field lower than 1 T. In this study, all the conducting In-free $Sn_xPb_{1-x}Te$ samples ($x \geq 0.05$) show the p-type conduction. PbTe ($x = 0$) that exhibits an insulating behavior at lower temperatures also shows the p-type conduction at high temperatures. This is in contrast to the previous literature, where a bulk PbTe crystal was reported to show a metallic conduction with n-type carrier [27]. We speculate that the widening of p-type conduction regime in our epitaxial thin films may come from the difference in the crystal defect formation which causes the deviation from charge neutrality.

Next, we examine the effect of In doping on $Sn_xPb_{1-x}Te$ whose $x$ ranges from 0.53 to 1.0, *i.e.* Sn-rich region. Figure 2(a) shows the In-doping dependence of $\rho_{xx}$ for five series of samples with different $x$ values ($x \geq 0.53$). While $\rho_{xx}$ in all the $y = 0$ samples are on the order of $10^{-4}$ to $10^{-3}$ $\Omega$cm as seen in Fig. 1(e), $\rho_{xx}$ increases by one or two orders of magnitude with In doping. A remarkable feature is that some samples with $x \geq 0.63$ show superconductivity as evidenced by a sharp drop in $\rho_{xx}$. The relation between carrier density and In composition is shown in Fig. 2(b). In the series with Sn composition $x = 1$ and 0.87, hole density $p$ increases with increasing In composition $y$. On the contrary, in $x = 0.63$ and 0.53 series, $p$ decreases for a small amount of In doping. Such a non-monotonic $y$-dependence of $p$ with varying $x$ can be understood by the valence skipping character of the In element [28]. In other words, In is incorporated as nominal $In^+$ (acceptor) in large $x$ region and nominal $In^{3+}$ (donor) in small $x$ and $y$ region. In Fig. 2(b), the samples that show the onset feature of superconductivity above 1.8 K are represented with filled circles. The superconductivity appears in samples whose $p$ value is approximately larger than $10^{20}$ cm$^{-3}$. Figure 2(c) shows the onset temperature of superconducting transition $T_c^{onset}$ as a function of $p$ for each $x$ series. Here, $T_c^{onset}$ is defined as the crossing point of linear extrapolation from normal conducting region and superconducting transition region (as exemplified by the inset of Fig. 2(c)). We can find the tendency that the sample with larger $p$ shows higher $T_c^{onset}$. The maximum $T_c^{onset}$ in our samples 3.73 K (($x$, $y$) = (0.63, 0.23)) is comparable with the value in a bulk crystal [24].

We examine the transport properties of the SPIT series from $x = 0.0$ to 0.42, i.e. Pb-rich region, where In is expected to act as a donor. The temperature dependence of $\rho_{xx}$ (Fig. 3(a)) shows a variety of transport characteristics from metallic to insulating behavior depending on $x$ and $y$ values, which contrasts with the metallic transport widely observed in the $x \geq 0.53$ series. With increasing $y$, the carrier density at $T = 2$ K shown in Fig. 3(b) represents the reduction of p-type carriers at $x = 0.30$, and furthermore a p-type to n-type transition at $x = 0.16$, both of which are consistent with the donor character of doped In as seen for $x = 0.63$ and 0.53 series in Fig. 2(b). The overall trend of the carrier type inversion is also clarified in the color maps of carrier density as functions of $x$ and $y$ (Fig. 3(c)). The insulating or semiconducting temperature dependence of $\rho_{xx}$ is explained by the reduction of carrier density that is associated with the carrier inversion. In particular, the sample series of $x = 0.23$ with $y \geq 0.09$ show a diverging behavior of $\rho_{xx}$ towards low temperatures, which suggests the trivial gap opening in bulk bands. On the other hand, the $x = 0.16$ series do not show such a resistivity divergence at the carrier type inversion but a metallic behavior below $T = 20$ K ($y = 0.09$, 0.17 and 0.23), representing the possibility of gap closing. In addition, the color map of mobility $\mu = 1/\rho_{xx}ne$ or $1/\rho_{xx}pe$ (Fig. 3(d)), where $e$ represents the elemental charge, shows a high mobility region centered at around $(x, y) = (0.16, 0.07)$ exceeding 5,000 cm$^2$V$^{-1}$cm$^{-1}$. The high carrier mobility suggests the presence of the semimetallic electronic states.

To discuss the possibility of polar semimetal states, the presence or absence of inversion symmetry is examined for the samples near the gap closing compositions by optical SHG experiment. Figures 4(a) shows the incident light polarization dependence of $p$-polarized SH intensity for the representative SPIT samples. The *p*-polarized SHG is clearly observed for *p*-polarized incident light in all the samples regardless of carrier type, which suggests the breaking of spatial inversion symmetry, namely the onset of the polarity along the out-of-plane of direction [111]. The observed large temperature and composition dependence of the SH intensity (Fig. 4(b)) indicates that the symmetry breaking originates from the electronic states in bulk region rather than surface/interface. In particular, SH intensities are apparently large in samples with $(x, y)$ = (0.16, 0.02) and (0.16, 0.09) that have large conductivity and high mobility (Fig. 3(d)), as compared with the insulating ones such as $(x, y)$ = (0, 0) and (0.23, 0.23). The breaking of spatial inversion symmetry near the gap closing that is concomitant with high charge mobility is thus reminiscent of the polar semimetal state such as Weyl semimetal.

In summary, we investigated the electronic states in $(Sn_xPb_{1-x})_{1-y}In_yTe$ thin films by electrical transport and optical SHG properties. By utilizing MBE thin film growth technique, we precisely tuned the chemical composition and examine the effect of In doping $y$ while varying the Sn/Pb composition ratio $x$. Superconductivity appears in the region of $(x \geq 0.63, y \geq 0.17)$, where the dopant of In acts as an accepter and increases the hole density. On the other hand, emergence of a semimetal state with high electron mobility is identified in the region centered at $(x, y)$ = (0.16, 0.07), where the n-type carrier is slightly doped by the donor character of In. Broken inversion symmetry in bulk states of the thin film is detected by optical SHG measurement for those high mobility samples, suggesting the possible emergence of polar semimetal states with bulk gap closing. The versatile electronic states in such telluride based topological materials will lead to the emergent phenomena such as topological superconductivity and anomalous Hall effect, in particular when they are proximitized with each other in a form of heterostructure [10,12,29].

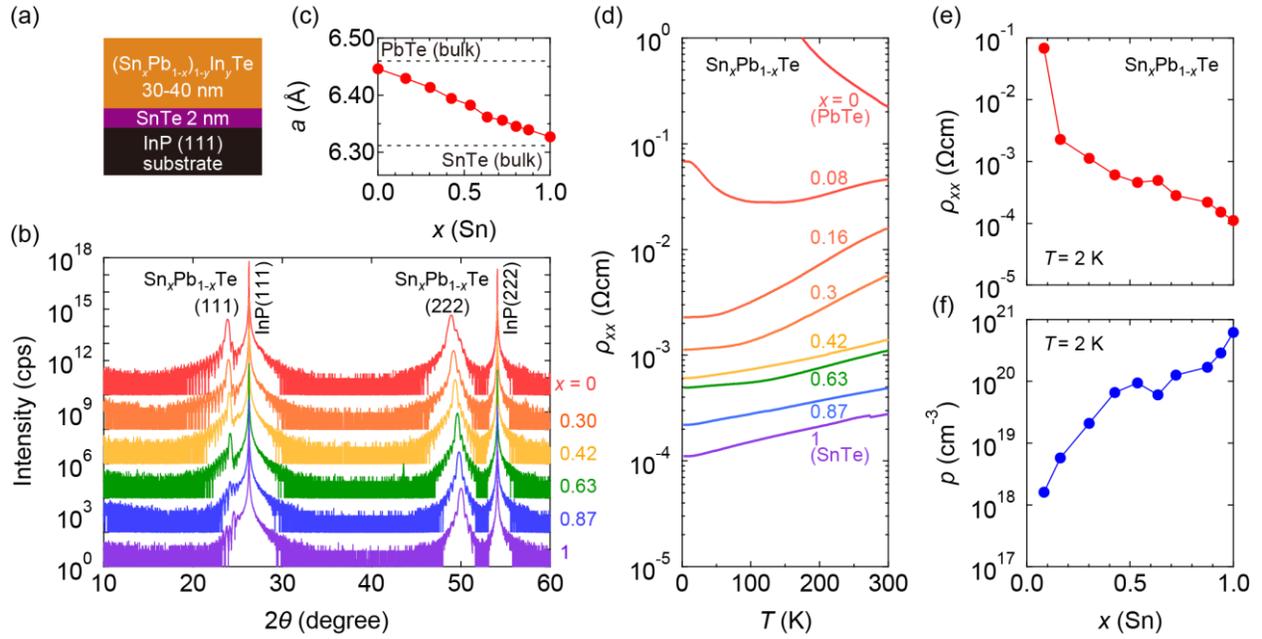

Figure 1 (a) Schematic of the sample structure. (b) X-ray diffraction (XRD) patterns in $2\theta$-$\omega$ scans for $Sn_xPb_{1-x}Te$ thin films with various $x$. (c) Cubic lattice constant evaluated from the (222) diffraction peak in Fig. 1(b). Dashed lines represent the bulk values for the lattice constant of PbTe and SnTe. (d) Temperature dependence of resistivity for $Sn_xPb_{1-x}Te$ thin films with several $x$. (e,f) Sn composition $x$ dependence of resistivity $\rho_{xx}$ (e) and hole carrier density $p$ (f) at $T = 2$ K.

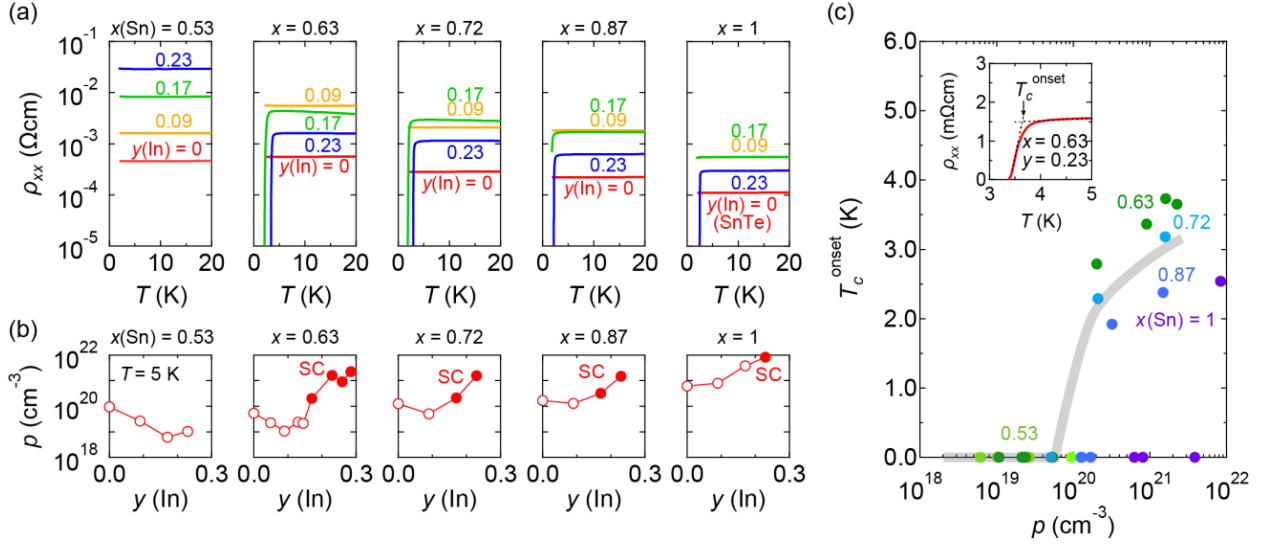

Figure 2 (a) Temperature dependence of longitudinal resistivity $\rho_{xx}$ for several $(Sn_xPb_{1-x})_{1-y}In_yTe$ samples with various $x$ ($x > 0.5$) and $y$ ($y \leq 0.23$) values. (b) $y$ dependence of hole density ($p$) at $T = 5$ K for series of samples with different $x$. Filled circles indicate the samples that shows superconductivity (SC). (c) $p$ dependence of onset temperature of superconductivity $T_c^{onset}$ for series of samples with different $x$. Inset shows the definition of $T_c^{onset}$ for the sample with $(x, y) = (0.63, 0.23)$.

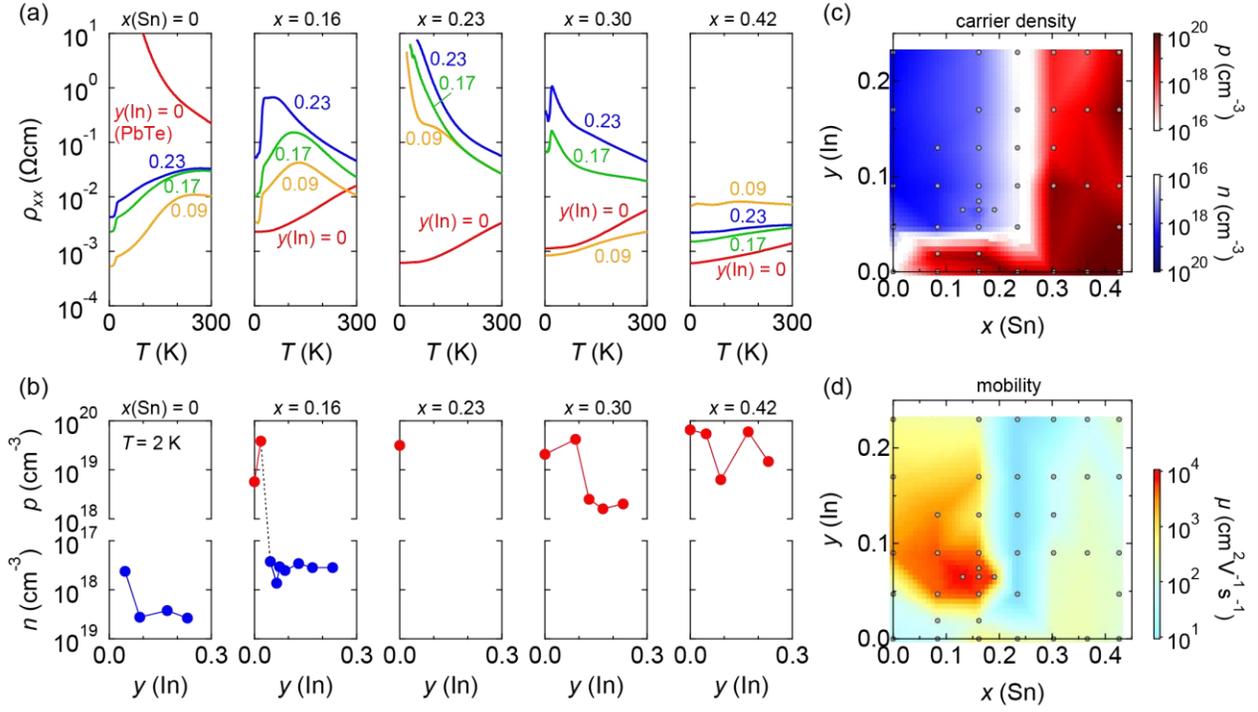

Figure 3 (a) Temperature dependence of longitudinal resistivity $\rho_{xx}$ for several $(Sn_xPb_{1-x})_{1-y}In_yTe$ samples with various $x$ ($x < 0.5$) and $y$ ($y \leq 0.23$) values. (b) $y$ dependence of electron (blue circles) and hole (red circles) density ($n$ and $p$, respectively) at $T = 2$ K for series of samples with different $x$. (c, d) Color maps of carrier density (c) and mobility (d) as functions of $x$ and $y$. The measured data points are overlaid on the color map with small gray circles.

Figure 4 (a) Incident light polarization dependence of *p*-polarized second harmonic (SH) intensity for six samples with various $x$(Sn) and $y$(In) values. (b) Temperature dependence of the SH intensity with $p_{in}$-$p_{out}$ geometry.

## ACKKNOWLEDGMENTS

We thank Tian Liang for fruitful discussion. This work was partly supported by the Japan Society for the Promotion of Science through JSPS/MEXT Grant-in-Aid for Scientific Research (Nos.18H01155 and 19J22547), and JST CREST (Nos. JPMJCR16F1, JPMJCR1874).
16

Ryutaro Yoshimi[1*], Makoto Masuko[2], Naoki Ogawa[1], Minoru Kawamura[1],

Atsushi Tsukazaki[3], Kei S. Takahashi[1], Masashi Kawasaki[1,2], and Yoshinori Tokura[1,2,4]

[1] *RIKEN Center for Emergent Matter Science (CEMS), Wako 351-0198, Japan.*

[2] *Department of Applied Physics and Quantum-Phase Electronics Center (QPEC),*

*University of Tokyo, Tokyo 113-8656, Japan*

[3] *Institute for Materials Research, Tohoku University, Sendai 980-8577, Japan*

[4] *Tokyo College, University of Tokyo, Tokyo 113-8656, Japan*

* Corresponding author: ryutaro.yoshimi@riken.jp

1. **X-ray diffraction for $x = 0.63$ and 0.87 series**

Figure S1 shows the X-ray diffraction data for $x = 0.63$ and 0.87 $(Sn_xPb_{1-x})_{1-y}In_yTe$ (SPIT) thin films. As shown in $2\theta$-$\omega$ scan (Fig. S1(a)), no secondary phase is discerned in both series. The reason why (111) diffraction peak is not clear in the $x = 0.87$ series compared with the 0.63 one is as follows. The peak intensity of (111) diffraction of rocksalt structure is in proportional to the difference of atomic scattering factors of cation and anion. Because the atomic scattering factors of Sn and Te are closer than those of Pb and Te, the zero-th order peak of (111) diffraction is weak in samples with large Sn compositions. This is the reason why we take (222) for evaluating cubic lattice constant in the main text. The $y$ dependence of cubic lattice constant $a$ for the both series is shown in Fig. S1(b). Although it is difficult to claim whether the $y$ dependence of $a$ follows the Vegard's law within the range of the present study ($y \leq 0.23$), $a$ tends to decreases with increasing $y$.

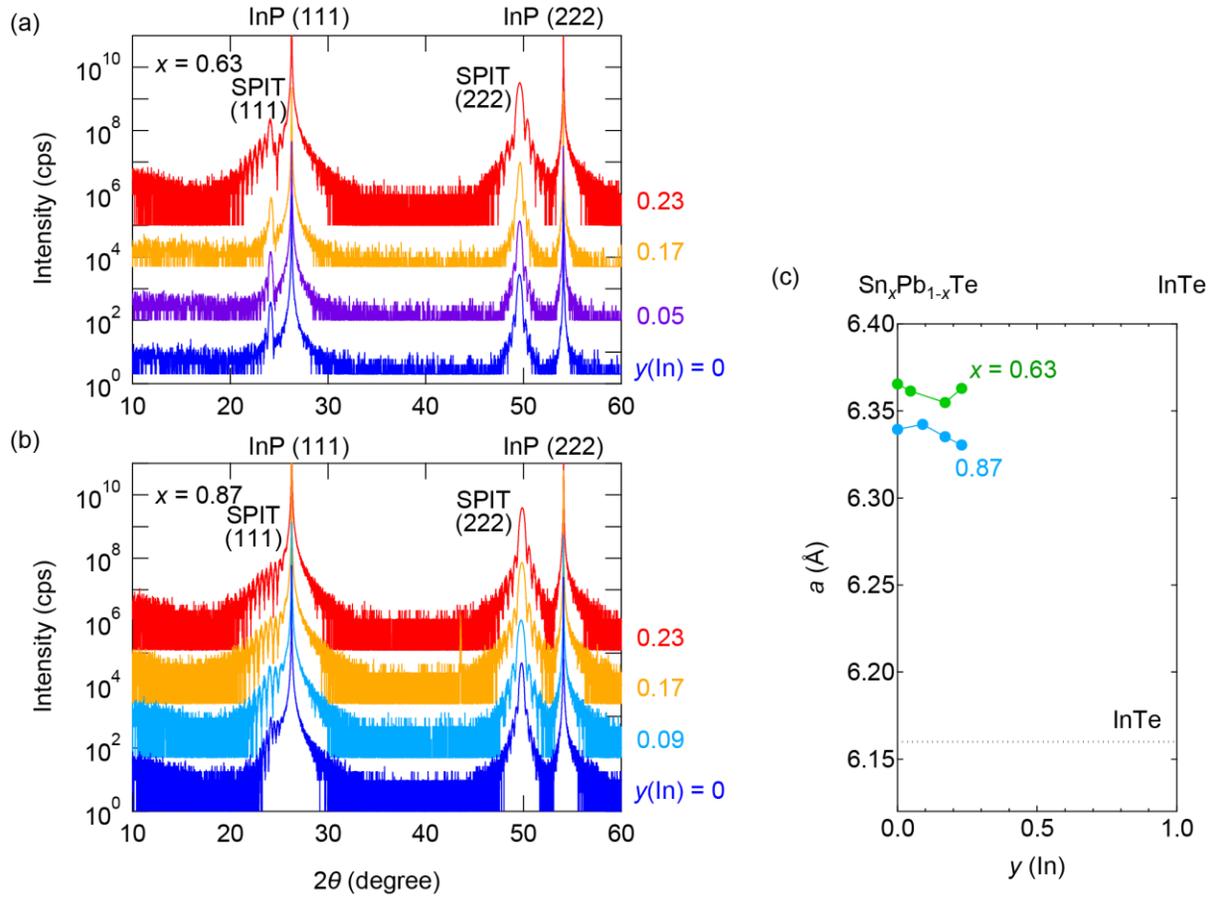

Fig. S1 (a), (b) X-ray diffraction patterns in $2\theta$-$\omega$ scans of $(Sn_xPb_{1-x})_{1-y}In_yTe$ thin films with different $y$(In) for $x$(Sn) = 0.62 (a) and 0.87 (b). (c) Indium doping dependence of cubic lattice constant $a$ evaluated from (222) diffraction peak for $x$ = 0.63 and 0.87 series. The broken line indicates the bulk value of the lattice constant for InTe.